\documentclass[grl]{agutex2015}

 \usepackage{graphicx}
\authorrunninghead{YORDANOVA ET AL.}

\titlerunninghead{Current sheets in the magnetosheath}

\begin{document}

%
%

\title{Electron scale structures and magnetic reconnection signatures in the turbulent magnetosheath}

\authors{E. Yordanova,\altaffilmark{1}
Z. V\"{o}r\"{o}s, \altaffilmark{2,3}
A. Varsani, \altaffilmark{2}
D. B. Graham, \altaffilmark{1}
C. Norgren, \altaffilmark{1}
Yu. V. Khotyaintsev, \altaffilmark{1}
A. Vaivads, \altaffilmark{1}
E. Eriksson, \altaffilmark{1}
R. Nakamura,\altaffilmark{2}
P.-A. Lindqvist, \altaffilmark{4}
G. Marklund, \altaffilmark{4}
R. E. Ergun,\altaffilmark{5}
W. Magnes,\altaffilmark{2}
W. Baumjohann, \altaffilmark{2}
D. Fischer, \altaffilmark{2}
F. Plaschke, \altaffilmark{2}
Y. Narita, \altaffilmark{2}
C.T. Russell, \altaffilmark{6}
R.J. Strangeway, \altaffilmark{6}
O. Le Contel, \altaffilmark{7}
C. Pollock, \altaffilmark{8}
R.B. Torbert, \altaffilmark{9}
B.J. Giles, \altaffilmark{8}
J.L. Burch, \altaffilmark{9}
L. A. Avanov \altaffilmark{8}
J. C. Dorelli \altaffilmark{8}
D. J. Gershman \altaffilmark{8,10}
W. R. Paterson \altaffilmark{8}
B. Lavraud \altaffilmark{11,12}
Y. Saito \altaffilmark{13}
}

{1}{Swedish Institute of Space Physics, Uppsala, Sweden.}
{2}{Space Research Institute, Austrian Academy of Sciences, Graz, Austria.}
{3}{Department of Geophysics and Space Sciences, E\"{o}tv\"{o}s University, Hungary.}
{4}{Royal Institute of Technology, Stockholm, Sweden.}
{5}{Laboratory of Atmospheric and Space Physics, University of Boulder, USA.}
{6}{University of California, Los Angeles, USA.}
{7}{Laboratoire de Physique des Plasmas, CNRS/Ecole Polytechnique/UPMC/Univ. Paris Sud/Obs. de Paris, Paris, Paris-Sud, France.}
{8}{NASA Goddard Space Flight Center, Greenbelt, MD, USA.}
{9}{Southwest Research Institute, San Antonio, USA.}
{10}{University of Maryland, College Park, MD, USA.}
{11}{IRAP, CNRS, Toulouse, France.}
{12}{ CNRS, UMR 5277, Toulouse, France}
{13}{JAXA, Japan.}



\keypoints{(Magnetic Reconnection, Turbulence)}

\begin{abstract}
Collisionless space plasma turbulence can generate reconnecting thin current sheets as suggested by recent results of numerical magnetohydrodynamic simulations. The MMS mission provides the first serious opportunity to check if small ion-electron-scale reconnection, generated by turbulence, resembles the reconnection events frequently observed in the magnetotail or at the magnetopause. Here we investigate field and particle observations obtained by the MMS fleet in the turbulent terrestrial magnetosheath behind quasi-parallel bow shock geometry.  We observe multiple small-scale current sheets during the event and present a detailed look of one of the detected structures. The emergence of thin current sheets can lead to electron scale structures where ions are demagnetized. Within the selected structure we see signatures of ion demagnetization, electron jets, electron heating and agyrotropy suggesting that MMS spacecraft observe reconnection at these scales.
\end{abstract}

\begin{article}
\section{Introduction}
The main goal of Magnetospheric Multiscale (MMS) mission is the multi-point study
of microphysics of magnetic reconnection (MR) targeting the structures within the electron diffusion region \citep{Burch15}. Additional science goals include the understanding of the physics of particle acceleration and the clarification of the role of plasma turbulence
in fast collisionless MR. On the other hand, high Reynolds number magnetohydrodynamic and PIC simulations show that
turbulence can also generate spatially intermittent, thin and reconnecting current sheets \citep{Greco08, Servidio09, Wan15}. The occurrence of MR in the turbulent terrestrial magnetosheath was also confirmed by Cluster measurements \citep{Retino07}. In turbulent space plasmas the ion-electron-scale current sheets are found to be associated with locally enhanced  heating and energy dissipation \citep{Osman12, Osman14, Chasapis15}. Although in collisionless plasmas only approximate measures of energy dissipation can be introduced \citep{Matthaeus15}, the generation of spatially intermittent current sheets indicates that the associated kinetic dissipation, in which MR can play a crucial role, is spatially inhomogeneous. Despite the highly localized dissipation the heating of the plasma can be significant. A recent experimental study based on Cluster data shows that turbulence generated thin proton-scale current sheets are ubiquitous in the magnetosheath downstream of a quasi-parallel bow shock \citep{Voros16}. This implies that turbulence may also generate numerous reconnecting current sheets which can be studied through high resolution field, plasma and particle measurements available from MMS.
Secondary MR sites can also occur at MR generated flux ropes or in turbulent reconnection exhausts \citep{Lapenta15}. The large number of turbulence generated or secondary MR sites may substantially increase the probability of MMS encounter by the electron diffusion region.
Global hybrid and fully kinetic simulations of the Earth's magnetosphere indicate that flux ropes and other plasma structures in the turbulent magnetosheath can also be generated by the interaction of the solar wind with the bow shock \citep{Karima14}.
In this paper we investigate a possible MR site in the turbulent terrestrial magnetosheath by detailed analysis of field, plasma and particle observation by MMS spacecraft. The different terms in the generalized Ohm's law are calculated and their relative contribution characterizing the ion and electron motion. Additionally, dimensionless proxies characterizing electron demagnetization and frozen flux violation are calculated from single point measurements \citep{Zenitani11,Aunai13,Scudder15}.


\section{Data and instrumentation}
The merged digital fluxgate (FGM) \citep{Russell14} and search coil (SCM) \citep{LeContel14} data was developed by using instrument frequency and timing models that were created during ground calibration. The inverse models were then applied to the respective inflight data. Data below 4Hz originates from FGM, data above from SCM and in the crossover region both data sets were used. The electric field data from EDP instrument is available with time resolution of 8 kHz \citep{Torbert14, Ergun14, Lindqvist14}. Ion and electron moments from FPI instrument \citep{Pollock16} have time resolution 150 ms and 30 ms, respectively.

\section{Event overview}
On November 30, 2015 between 00:21 and 00:26 UT the MMS spacecraft were situated in the compressed turbulent magnetosheath, downstream of a quasi-parallel bow shock. At the same time, the solar wind monitors (OMNI database) observed an extended high-density compressional region at the leading edge of a high-speed stream, associated with a significant geomagnetic response (not shown). The overview plot (Figure 1) shows the observed field and plasma parameters between 00:26:03 and 00:26:18 UT. It is demonstrated here that this 15 sec long interval contains a flux rope and its interacting boundary/region comprising discontinuities, narrow current structures, and magnetic reconnection. These are the typical structures seen in simulations of plasma turbulence \citep{Greco08, Servidio09, Wan15}.
The subplots $1a-d$ show the total magnetic field B$_{t_k}$ and  magnetic components B$_{x_k}$, B$_{y_k}$ and B$_{z_k}$ in spacecraft reference frame. Indices $k$ refer to MMS spacecraft. The so-called partial variance of increments, $PVI_{ij}$, calculated between spacecraft pairs $i,j$ ($i,j$=1--4 are the number of MMS spacecraft), are often used in the studies of plasma turbulence to detect discontinuities or current sheets. $PVI_{ij}$ are depicted in the subplot $1e$, and are defined through:
\begin{equation}
PVI_{ij}(t)=\sqrt{\frac{\mid{\Delta \textbf{\textit{B}}_{ij}(t)\mid}^2}{\big \langle{\mid{\Delta \textbf{\textit{B}}_{ij}}\mid ^2}\big \rangle}},
\label{eq1}
\end{equation}
The latitude $\theta_2$ and longitude $\phi_2$ of magnetic field vector orientation for MMS~2 is shown in subplot $1f$. Subplot $1g$ contains the pressures (total, dynamic, magnetic, ion thermal and electron thermal). Here the different pressure terms are shown with the same color for each spacecraft. The magnitudes of ion and electron speeds, \textbf{V}i$_k$ and \textbf{V}e$_k$ are shown in subplots $1h$ and $1i$, respectively. The magnitudes of electric field \textbf{E$_k$} and the magnitudes of current densities \textbf{J$_k$} are shown in subplots $1j$ and $1k$, respectively. \textbf{J$_k$}'s are calculated for each spacecraft from plasma measurements through \textbf{J}$_k$ = $Nq$(\textbf{V}i$_k$-\textbf{V}e$_k$), where $N$ is the plasma density and $q$ is the charge of particles. The thick magenta line in subplot $1k$ corresponds to the magnitude of the current density \textbf{J$_{curl}$}, estimated in the tetrahedron barycenter by using the curlometer technique \citep{Dunlop02}.

There exist two different physical regions which can be identified in Figure 1. A twisted flux rope extends roughly from 00:26:10 UT to the end of the time interval. It can be identified on the basis of the slow rotation and sign-change of the magnetic field, seen in $Bz_k$ (subplot $1d$), changing from -32 nT (minimum) to +8 nT (maximum). The slow rotation is also seen in $\theta_2$. Other signatures of the flux rope include the maxima of $Bt_k$ (subplot $1a$) and total pressure $P_{tot}$ (subplot $1g$) between 00:26:11 and 00:26:13 UT. Although the ion P$_{therm}$ is higher then $P_{mag}$, the profile of $P_{tot}$ having maximum near the rope axis \citep{Zaqar14} is determined by $P_{mag}$.

At the left border of the flux rope (roughly between 00:26:05 and 00:26:10 UT) we observe a distinct feature in all parameters. Further in the paper we will refer to it as the region of interest. Within this region the differences between magnetic field values (subplots $1a-d$) become larger, indicating increased magnetic gradients. $PVI_{ij}$ show the occurrence of two discontinuities (subplot $1e$), where also the orientations of magnetic vectors ($\theta$ and  $\phi$ in subplot $1f$) exhibit sudden changes. At the same time, there exist significant changes and narrow peaks in $Ve_k$, $E_k$ and $J_k$, while the $Vi_k$ variations are much smaller, indicating the occurrence of differential motion between ions and electrons at narrow structures. The electron inertial length in this region is  $\sim 0.7$ km and the Doppler shifted frequency associated with this scale corresponds to about 26 Hz. These structures are narrower than the inter-spacecraft separation ($\sim$ 10 km), therefore the curlometer cannot detect them (the magenta curve of $J_{curl}$ is much smoother in subplot $1k$).
The narrow peaks in \textbf{V}e$_k$, \textbf{E$_k$} and \textbf{J$_k$} between  00:26:08.5 and 00:26:10 UT are subsequently seen by all spacecraft, therefore representing  real spatial structures.

\section{The event in a new coordinate system}
To better understand the event presented in Figure 1 the physical variables were rotated to the field-aligned coordinate system, in which X: $\textbf{B}$, Y: \textbf{E}$\times $\textbf{B}  and Z: \textbf{B}$\times $\textbf{E}$\times $\textbf{B}. We have chosen a rotation matrix at the instant of electron speed maximum before 00:26:10 UT in Figure 2 (subplot $2d$), which served as a global coordinate system for the whole event. In this coordinate system the largest variations of the magnetic field occur at the border of the flux rope in $B_x$ and $B_z$ components, while  $B_y$ is changing slowly (subplots $2a-c$). The flux rope interval after 00:26:10 UT is characterized by a slow rotation of the magnetic field. The electron speed components (subplots $2d-f$) show occurrence of jets at the border, while the ion speed increase is smaller and the variation is smoother (subplots $2g-j$). Similar electron jets have been observed at the magnetopause [$Khotyaintsev$, $et$ $al.$, 2016, this issue]. The ion and electron Alfv\'en speeds vary between spacecraft from 115 to 125 km/s . Electron and proton  parallel and perpendicular temperatures are shown in subplots $2i$ and $k$. $Te_{\parallel}$ (subplot $2i$) shows two peaks associated with temperature anisotropy and parallel electron heating at the left and right borders of the region between 00:26:05 and 00:26:11 UT in all spacecraft. Similar increases have been observed by recent MMS measurements at the magnetopause near the diffusion region and have been interpreted as evidence for a potential reconnection exhaust \citep{Graham16, Lavraud16}. The ion temperature anisotropy however, is absent within this region (subplot $2k$).  The slight increase of ion plasma density (subplot $2l$) together with the increase of magnetic field (subplot $2a$) and total/magnetic pressure (Figure $1g$) between 00:26:09 and 00:26:11 UT indicates that this is a compressional region.
The fluctuations and temperature anisotropies after 00:26:11 UT are associated with the flux rope again.

\section{Generalized Ohm's law terms}
In collisionless plasmas magnetic reconnection represents a multi-scale process where characteristic reconnection structures over different scales can be observed. It is described by the generalized Ohm's law been written in terms of the electric field \textbf{E} \citep{Khotyaintsev06}:

\begin{equation}
E+ \mathbf{V_i}\times \mathbf{B}= \frac{\mathbf{J} \times \mathbf{B}}{ne} +\frac{\nabla \cdot \mathbf{P}_e}{ne}
\label{eq2}
\end{equation}
where \textbf{V} is the plasma bulk flow speed, \textbf{J} is the current density, \textbf{P} $_e$ is the electron pressure tensor, \textbf{E}  is the electric field, \textbf{B} is the magnetic field, $m_e$ is the electron mass, $n$ is the number density, and $e$ is the proton charge.
The $z$ (out-of-plane) components of the terms in the generalized Ohm's law and their relative strength indicate if the spacecraft are crossing the ion or electron diffusion regions \citep{Nakamura16}. The different terms in the Ohm's law are plotted in Figure 3 $a-e$. The (\textbf{V$_e $}$\times $\textbf{B})$_z$ is small (subplot $3a$), however, the electric field in the electron frame (\textbf{E}+\textbf{V}$_e$ $\times $\textbf{B})$_z$ is large around 00:26:10 UT (subplot $3b$).  The Hall term (\textbf{J}$\times$\textbf{B}/ne)$_z$ in subplot $3c$ indicates that (\textbf{V}$_e$ $\times$ \textbf{B})$_z$ $\gg$ (\textbf{V}$_i$ $\times$ \textbf{B})$_z$ and the differential motion of electrons and ions leads to significant Hall terms. The \textbf{E}$\cdot$\textbf{J}  reaching large values in the region of interest (subplot $3d$) indicates that electromagnetic energy is converted to thermal and kinetic energies. 

Additionally, Figure $3f$ shows the $\sqrt{Q}$ parameter introduced by \citep{Swisdak16} representing a measure of gyrotropy of the electron pressure tensor.  It is defined as following:
\begin{equation}
Q=1-\frac{4I_2}{(I_1-P_{\| })(I_1+3P_{\| })},
\label{eq3}
\end{equation}
where $I_1=P_{xx}+P_{yy}+P_{zz}$, $I_2=P_{xx}P_{yy}+P_{xx}P_{zz}+P_{yy}P_{zz} - (P_{xy}P_{yx}+P_{xz}P_{zx}+P_{yz}P_{zy})$, and $P_{\|}$=\textbf{\^{b}}$\cdot$\textbf{P}$\cdot$\textbf{\^{b} }. For gyrotropic tensors $Q$=0, while maximal agyrotropy is reached at $Q$=1.
The variations due to electron pressure tensor, ($\nabla \cdot$  \textbf{P}$_e$/ne)$_z$    (subplot $3e$) are also elevated when the electric field in the electron frame is (subplot $3b$) is high. Similar behavior is observed at the magnetopause [\textit{Norgren et al.}, 2016, this issue].  Finally, according to PIC simulations \citep{Swisdak16} the parameter $\sqrt{Q}$ reaching values about $0.05$ indicates significant agyrotropy, which occurs near the separatrices or reconnection X-lines. This is most pronounced at about 00:26:09.5 UT for MMS~1 (black peak in subplot $3f$), note however that $\sqrt{Q}$ is enhanced within the whole interaction region.

\section{Particle distributions}
Figure 4, represents the plasma observations by FPI ion and electron instrument on MMS~1. The top four stack plot are energy spectrograms, which respectively show the ion distribution perpendicular to the local magnetic field, and also electron distribution parallel, perpendicular and anti-parallel to the magnetic field. From the ion energy spectrogram, it is evident that at 00:26:00 UT the ion population had an energy centered at $\sim 750$ eV. Since then the flux of ions showed some variations, however the center of energy remained the same.  The first clear change occurs at $\sim$ 00:26:07.5 UT around the time that total magnetic field reached its minimum value, where the flux of ions also increased. Then at $\sim$ 00:26:09.6 UT, a distinct colder population with energies centered at $\sim$ 150 eV emerged, whilst a lower flux population was also centered at $\sim 400$ eV. The colder magnetosheath ions are observed until $\sim$ 00:26:10.5 UT where a higher energy population, narrowly distributed around $\sim 500$ eV, appeared. The latter is the dominant population until 00:26:14.2 UT, when ions separated in two distinct populations, one centered at $\sim 300$ eV and the other at $1$ KeV. This trend continued until the end of the period at 00:26:20 UT, where the ions had one population with $\sim 700-800$ eV energy.

The energy spectrogram of the electrons shows that at the start of the period, they were mainly bistreaming, which continued until $\sim$ 00:26:03.4 UT when the distribution became rather isotropic. At $\sim$ 00:26:07.1 UT, the distribution turned to bistreamig for a short period (about $\sim 0.5$ second) before the minima of the total magnetic field. The population was again isotropic until $\sim$ 00:26:09.6 UT when the magnetic field $B_{z} = 0$. At this time, the population was predominantly anti-field aligned, resulting in the velocity of electrons reaching $\sim 600$ km/s purely in that direction. This narrow region quickly passed by MMS~1, and only $0.2$ second later, the electrons were observed to be moving mainly perpendicular to the magnetic field with speed of $\sim 550$ km/s, which lasted $\sim 0.1$ second. In this short interval the center of energy for the electrons quickly rose from $\sim 100$ eV to $\sim 150$ eV and then back to $\sim 100$ eV. This time interval, which is marked by the rectangle in Figure $4$, is when the values of $\sqrt{Q}$ reached their maximum, representing a non-gyrotropy in electron distribution. In addition to that, this signature is also accompanied with appearance of colder ion population as mentioned above, and therefore is of particular interest. After this time, the perpendicular electrons show variations at 00:26:10.6, 11.9, 12.9, 14.1 and 16.1 UT, whilst the parallel and antiparallel stays relatively equal. However, between 00:26:09.6 and 00:26:09.8 UT, it was the only time interval that the  maximum velocity and a clear increase in the flux of electrons were observed.

The bottom two rows of Figure 4 show the Velocity Distribution Function (VDF) of electrons and ions, respectively, for a snapshot at 00:26:09.710 and 00:26:09.800 UT. Each row contains two panels which show the cuts of VDFs in \textbf{V}$_{\parallel}$--\textbf{V}$_{\perp_1}$ and  \textbf{V}$_{\perp_1}$--\textbf{V}$_{\perp_2}$,  where \textbf{V}$_{\parallel}$ represents the velocity along the magnetic field orientation, \textbf{V}$_{\perp_1}$ and \textbf{V}$_{\perp_2}$ respectively along  (\textbf{E}$\times $\textbf{B}) and \textbf{B}$\times$(\textbf{E}$\times $\textbf{B}) directions.
The \textbf{V}$_{\parallel}$--\textbf{V}$_{\perp_1}$ plot for the electrons shows that, while the lowest energy population are approximately isotropic, there is also a population which were purely moving in positive (\textbf{E}$\times $\textbf{B})  direction. The former population as also mentioned above is $\sim 100$ eV and the later $\sim 150$ eV. The simultaneous observations of ions however show that the main population ($\sim 400$ eV) was  anti-field aligned,  whilst there was no clear (\textbf{E}$\times $\textbf{B}) drifted population. 

In  the \textbf{V}$_{\perp_1}$--\textbf{V}$_{\perp_2}$ plot for the electrons, the population with lowest energy ($<100$ eV) is gyrotropic, but the higher energy ($\sim 150$ eV) population show a clear non-gyrotropy with the electrons being shifted in positive (\textbf{E}$\times $\textbf{B})  direction. In a (\textbf{E}$\times $\textbf{B}) drifted distribution in plasma, it is expected that the lower energy particles are more effected due to the relative velocity drift. However for this case, the fact that the lower energy population is drifted less than the higher energy population, it may suggest that the latter is a distinct population. The simultaneous VDF for ions shows that the highest flux ions were predominantly moving in negative direction of \textbf{B}$\times$(\textbf{E}$\times $\textbf{B}) directions. This population is the same cold ions centre at $\sim 150$ eV, which emerged at the time of the rotation of $B_{z}$ component in the magnetic field. Also the lower flux ions with energy centred at $\sim 400$ eV, are approximately gyrotropic at this time.

\section{Discussion and summary}

The appearance of the two distinct cold and hot populations of ions leading to a non-maxwellian distribution, suggests that these observations have taken place where ions were demagnetized. (e.g. \citep{Dai15,Zhou09}). This idea is supported by the non-gyrotropic shape of the ion distributions in VDF plots, where an asymmetric reconnection (e.g. \citep{Lee14}) can lead to mixing up distinct sources of plasma with different energies. The separation of two populations in \textbf{V}$_{\perp_1}$--\textbf{V}$_{\perp_2}$ plane along is also consistent with previous observations of ion diffusion region by \citep{Dai15} using THEMIS spacecraft.

At electron scale, the narrow region with excessive anti-field aligned electron jet, followed by a jet along (\textbf{E}$\times $\textbf{B}) direction in the interval where E field reached its maximum, suggests a possible passage of the spacecraft near the X-line.  Similar signatures of diffused ions followed by electron outflow with electrons frozen-in on the reconnected field line have been observed in the separatrix region in simulation of asymmetric reconnection at the magnetopause [\textit{Khotyaintsev et al.}, 2016, this issue]. There is also an increase in the agyrotropy parameter Q for the electrons, which is expected to be seen around the electron diffusion region, however the observations are not accompanied with a crescent shaped distribution of electrons in \textbf{V}$_{\perp_1}$--\textbf{V}$_{\perp_2}$ plane (e.g \citep{Hesse99,Hesse11}). Note that this crescent shape is clearer for magnetopause reconnection where the two sources of plasma have clear energy differences, whilst in the magnetosheath it may not be the case. Overall, the particle data here suggest that most of the observations are near the X-line inside  ion diffusion region. The spacecraft does not clearly enter the electron diffusion region, however the non-diagonal elements of electron pressure tensor increase significantly as MMS~1 probably crossed the separatrix region. This is consistent with simulations on the spatial dimensions of the electron diffusion region (e.g. \citep{Nakamura16,Swisdak16}).

In summary, the main motivation for this study is to show that coherent structures such as flux ropes, current sheets, reconnection associated multi-scale structures, can be observed over proton and electron scales in the turbulent magnetosheath by MMS spacecraft. The analyzed time interval comprised a flux rope with slightly rotating magnetic field with compressions, discontinuities, current sheets, electron and ion scale ($\sim$ 0.5-30 km) structures developing at its border. In this region of interest, the four MMS spacecraft observed: 1) strong electron scale currents; 2) significant $z$ components of the electric field in the electron frame (\textbf{E}+\textbf{V}$_e$$\times $\textbf{B})$_z$  and the Hall term (\textbf{J}$\times$\textbf{B}/ne)$_z$; 3) signature of demagnetized ions and ion Alfv\'en outflow; 4) fast electron jets; 5) electron heating; 6)\textbf{E}$\cdot$\textbf{J} up to $\sim$ 70 $nW/m^3 $ at narrow peaks indicating that the electromagnetic energy is converted and dissipated; and 6) electron pressure agyrotropy. These features suggest that MMS presumably observes MR site within electron scale current sheets in the turbulent magnetosheath plasma.

This study, complementing and further developing the ideas about turbulence generated structures in the magnetosheath \citep{Retino07,Chasapis15,Voros16}  suggests that electron scale structures and reconnecting current sheets may occur not only at the large-scale boundaries, such as the magnetopause or magnetotail current sheet, but also in turbulent collisionless plasmas. We believe that these findings might encourage more thorough investigations of turbulence generated structures by using the high resolution measurements of MMS.


%
%
%
%
%
%
%

\begin{acknowledgments}
E.Y. and Z. V. research leading to these results has received funding from the European Community's Seventh Framework Programme ([$7/2007-2013$] under grant agreement $n^\circ  313038$/STORM. Z.V. was supported by the Austrian Fond zur F\"orderungur der wissenschaftlichen Forschung (project P24740-N27). IRAP contribution and the French involvment to (SCM) on MMS are supported by CNES and CNRS. Y.K. and D.B.G. were supported by the Swedish National Space Board, grants 139/12 and 175/15.
\end{acknowledgments}

\altaffiltext{1}{Swedish Institute of Space Physics, Uppsala, Sweden.}
\altaffiltext{2}{Space Research Institute, Austrian Academy of Sciences, Graz, Austria.}
\altaffiltext{3}{Department of Geophysics and Space Sciences, E\"{o}tv\"{o}s University, Hungary.}
\altaffiltext{4}{Royal Institute of Technology, Stockholm, Sweden.}
\altaffiltext{5}{Laboratory of Atmospheric and Space Physics, University of Boulder, USA.}
\altaffiltext{6}{University of California, Los Angeles, USA.}
\altaffiltext{7}{Laboratoire de Physique des Plasmas, CNRS/Ecole Polytechnique/UPMC/Univ. Paris Sud/Obs. de Paris, Paris, Paris-Sud, France.}
\altaffiltext{8}{NASA Goddard Space Flight Center, Greenbelt, MD, USA.}
\altaffiltext{9}{Southwest Research Institute, San Antonio, USA.}
\altaffiltext{10}{University of Maryland, College Park, MD, USA.}
\altaffiltext{11}{IRAP, CNRS, Toulouse, France.}
\altaffiltext{12}{ CNRS, UMR 5277, Toulouse, France}
\altaffiltext{13}{JAXA, Japan.}

%
%
\end{article}
%
%
%
%
%
%

\begin{figure}
\includegraphics[scale=0.75, bb=0 -50 0 600 ]{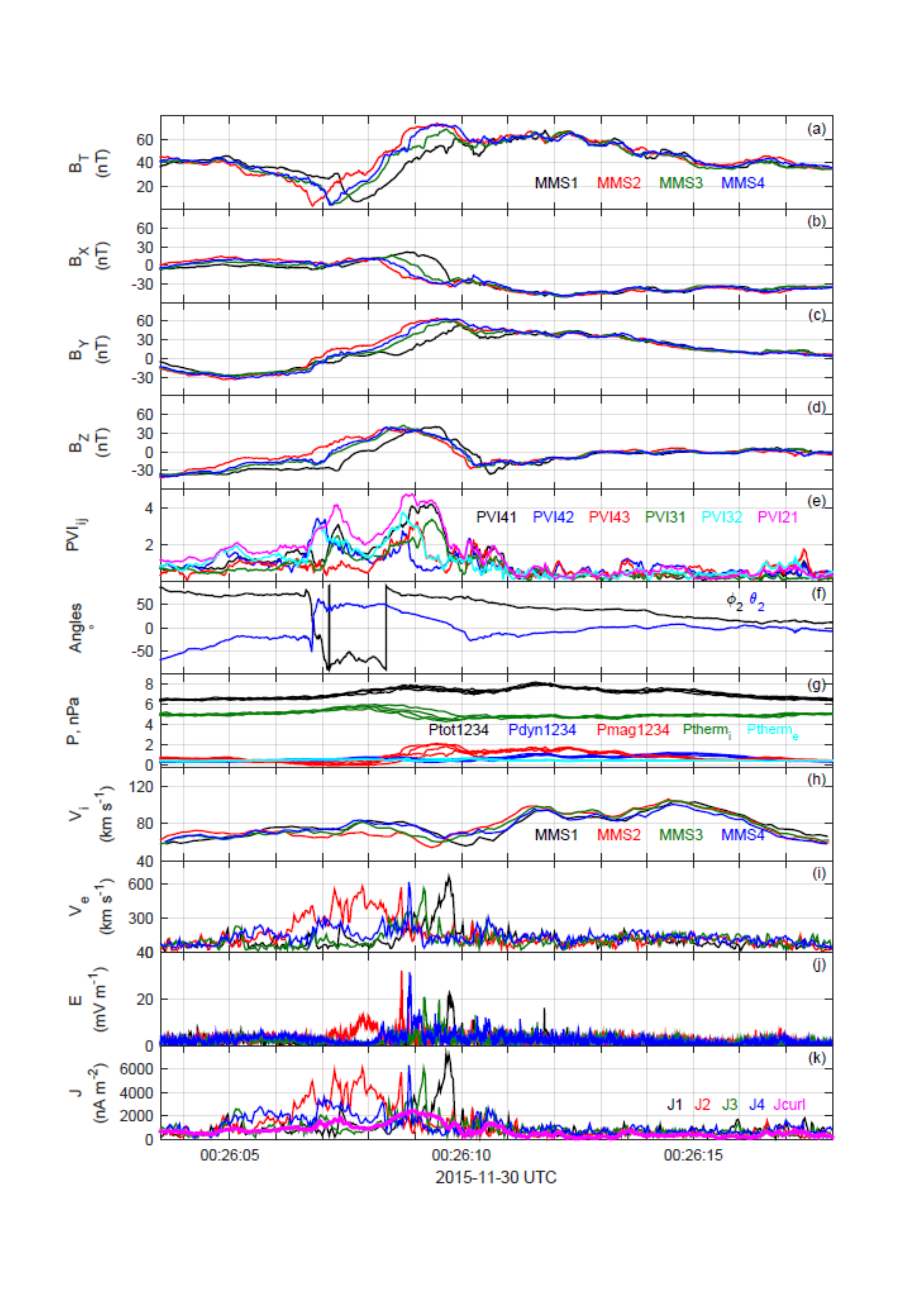}
\caption{Field and plasma parameters in spacecraft coordinates: a) Magnetic field magnitude for the four MMS spacecraft (color coded); b)-d) magnetic field components; e) $PVI_{ij}$ from pairs of spacecraft (i,j=MMS 1-4); f) the elevation and azimuthal angle of the magnetic field for MMS 2; g) Pressure for all spacecraft: total (black), dynamical (blue), magnetic (red), ion thermal (green), and electron thermal (cyan); h) ion velocity for all spacecraft; i) electron velocity; j) electric field magnitude; and k) electric current from plasma for each spacecraft and the current from curlometer (magenta).}
\label{fig1}
 \end{figure}

\begin{figure}
\noindent\includegraphics[scale=0.75, bb=0 -50 0 600 ]{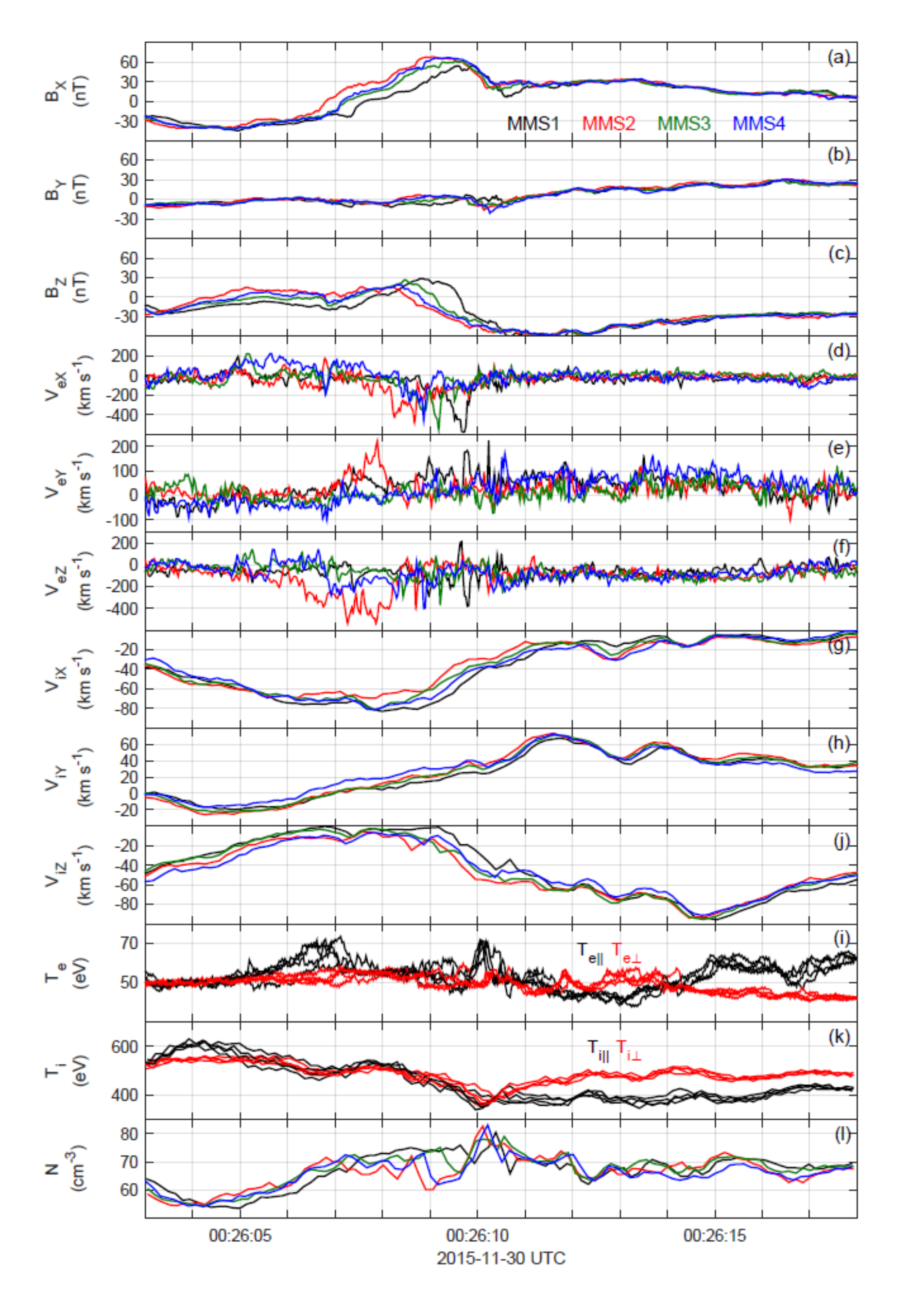}
\caption{Field and plasma parameters in the rotated coordinate system: a)–c) magnetic field components for all spacecraft; d)-f) electron speed components; g)-j) ion speed components; i)-k) parallel (black) and perpendicular (red) to the background magnetic field electron and ion temperature for all spacecraft, respectively; and l) ion density.}
 \label{fig2}
\end{figure}

 \begin{figure}
\noindent\includegraphics[scale=0.75, bb=0 -50 0 600 ]{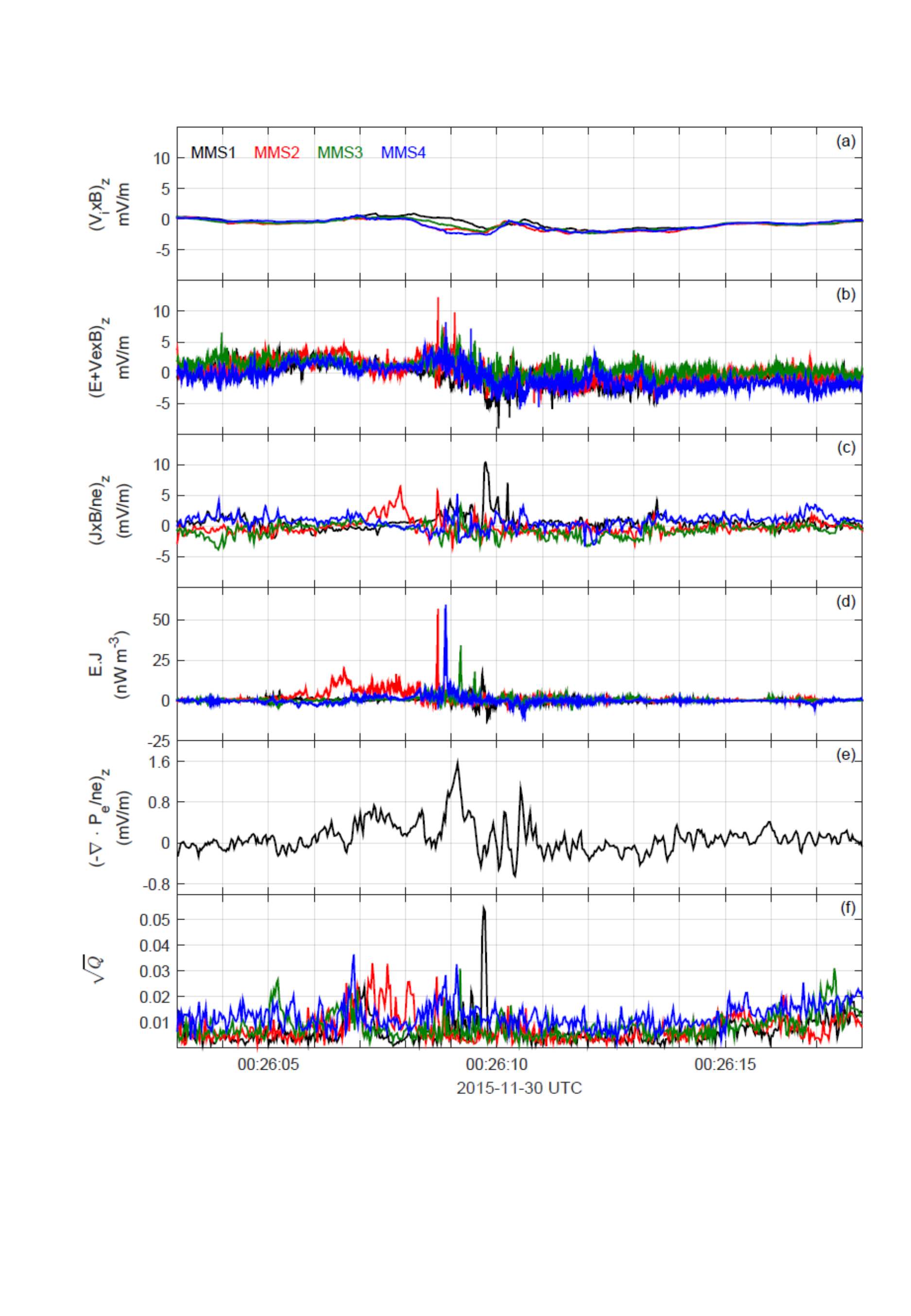}
\caption{Generalized Ohm's law terms: a) ion convection component in Z-direction; b) z- component of the electric field in electron frame; c) Hall term component in Z-direction; d) E.J dissipation; e) electron pressure term in Z; and f) agyrotropy parameter.}
 \label{fig3}
 \end{figure}

\begin{figure}
\noindent\includegraphics[scale=0.75, bb=0 -50 0 600 ]{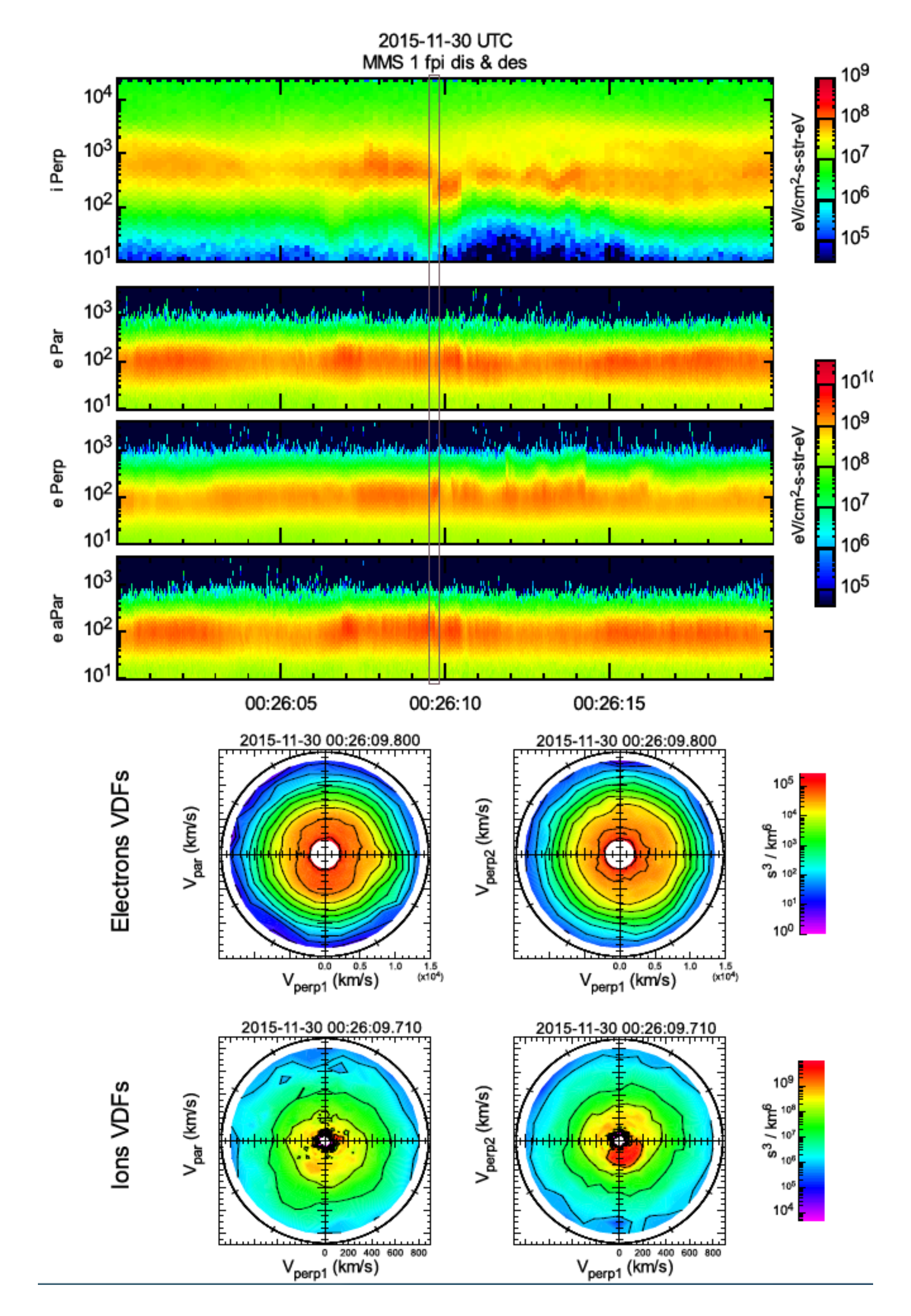}
\caption{ Energy spectrograms of particle distributions perpendicular and parallel to the local magnetic field for ions (upper two horizontal  panels), and for electrons (bottom two horizontal panels; Velocity Distribution Function (VDF) parallel and perpendicular to the magnetoc field of electrons (upper row of diagrams) and ions (bottom row of diagrams) of snapshots taken at the beginning and the end of the interval indicated by the narrow brown box in the spectrograms.}
 \label{fig4}
 \end{figure}

%


\end{document}